\documentclass[12pt]{article}

\usepackage{amssymb,amsmath,mathrsfs,epstopdf,slashed,color}
\usepackage{tikz}
\usetikzlibrary{matrix}
\usetikzlibrary{positioning}
\usepackage{ifpdf}
\ifpdf
\usepackage{graphicx}
\usepackage{placeins}
\usepackage{hyperref}
\usepackage{cite}
\else
\usepackage[dvipdfmx]{graphicx}
\usepackage[dvipdfmx]{hyperref}
\usepackage[numbers]{natbib}
\usepackage{hypernat}
\fi

\allowdisplaybreaks[1]

\makeatletter

\@addtoreset{equation}{section}
\makeatother

\DeclareMathOperator*{\tr}{tr}
\DeclareMathOperator*{\Tr}{Tr}

\begin{document}

\begin{titlepage}

\setcounter{page}{0}
  
\begin{flushright}
 \small
 \normalsize
\end{flushright}

\vskip 3cm
\begin{center}

{\Large \bf The Chern-Simons Number as a Dynamical Variable}  

\vskip 2cm
  
{\large S.-H. Henry Tye${}^{1,2}$ and Sam S.C. Wong${}^{1}$}
 
 \vskip 0.6cm

 ${}^1$ Department of Physics and Jockey Club Institute for Advanced Study,
 
  Hong Kong University of Science and Technology, Hong Kong\\
 ${}^2$ Laboratory for Elementary-Particle Physics, Cornell University, Ithaca, NY 14853, USA

 \vskip 0.4cm

Email: \href{mailto:iastye@ust.hk, scswong@ust.hk}{iastye at ust.hk, scswong at ust.hk}

\vskip 1.0cm
  
\abstract{\normalsize In the standard electroweak theory that describes nature, the Chern-Simons number associated with the vacua as well as the unstable sphaleron solutions play a crucial role in the baryon number violating processes.  We recall why the Chern-Simons number should be generalized from a set of discrete values to a dynamical (quantum) variable. Via the construction of an appropriate Hopf invariant and the winding number, we  discuss how the geometric information in the gauge fields is also captured in the Higgs field. We then discuss the choice of the Hopf variable in relation to the Chern-Simons variable.
 }
  
\vspace{1cm}
\begin{flushleft}
 \today
\end{flushleft}

 \vspace{3mm}
 
 {to appear in {\it Annals of Mathematical Sciences and Applications}}
\end{center}
\end{titlepage}

\setcounter{page}{1}
\setcounter{footnote}{0}

\section{Introduction}

Since the discovery of the Higgs Boson in 2012 \cite{Aad2012a,Chatrchyan2012a}, the standard theory of electromagnetic and weak interactions, namely the $SU(2) \times U(1)$ theory, is essentially established. This electroweak theory is in full agreement with all known observations. Furthermore, there is not the slightest  indication from available data that, except for the neutrino sector, this theory needs any extension or modification beyond its present form. However, features of its non-perturbative properties involving interesting geometry remain to be tested experimentally. 

Due to the non-Abelian nature of the $SU(2)$ (Yang-Mills) gauge fields, the vacuum structure of the theory is non-trivial. 
Following the construction of the instanton solution in 4-dimensional Euclidean spacetime with non-trivial topological Chern-Pontryagin (CP) index \cite{Belavin1975}, 'tHooft pointed out that a non-zero CP number leads to a change of the vacuum state together with a change of the baryon (or atomic) number $B$ and a change of the lepton number $L$ \cite{Hooft1976,Hooft1976a}.  (Recall that each of the 3 quarks in a proton or a neutron has baryon number $B=1/3$ and an electron has one unit of lepton number.) This non-conservation of $(B+L)$ in nature is expected to have deep implications on the matter-anti-matter asymmetry of our universe \cite{Kuzmin1985,Shaposhnikov1987,Fukugita1986f}, a possibility that has been extensively studied \cite{Cohen1993,Riotto1999a}. 

Typical $(B+L)$-violating processes take place via tunneling, so it is understood that the CP number should be generalized to a continuous variable that is both dynamical and quantum. However, attempts to carry out this construction for pure gauge theory \cite{Bitar1978} faces challenges like the size scaling property (thus requiring a cut-off) and Euclidean nature of instantons as well as the running of the gauge coupling. 

\begin{figure}[h]
 \begin{center}
 \label{figperiodic}
  \includegraphics[scale=.7]{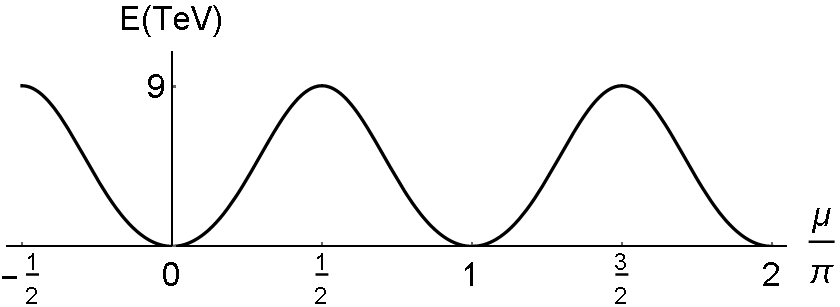}
  \caption{The periodic potential $V(\mu)$ as a function of $\mu/\pi$. Here we have chosen $\mu/\pi=0$ as the reference vacuum. The vacua are labelled by integer values while a barrier height, corresponding to the sphaleron static solution of $E_{sph}=9.0$ TeV, is at a half-integer value of $\mu/\pi$.}  
 \end{center}
\end{figure}

Fortunately in the electroweak theory, there is a complex scalar doublet field coupled to the $SU(2)$ gauge fields. 
Under spontaneous symmetry breaking, this Higgs field acquires a vacuum expectation value $v=246$ GeV that provides a mass/length scale to the theory and the coupling stays small.  A non-trivial static solution of the electroweak theory (in Minkowski spacetime) can be constructed. This static solution, namely the sphaleron, is unstable \cite{Manton1983,Klinkhamer1984}. It corresponds to the peak of the potential barrier that separates the vacua, as shown in FIG \ref{figperiodic}, where the height of the barrier (i.e., the sphaleron energy) is $9.0$ TeV and the resulting potential $V(\mu)$ is periodic. In FIG 1, the minima are vacua with Chern-Simons (CS) value $\mu/\pi$ at integer values while the peaks have $\mu/\pi$ at half-integer values.

It is well known that the CS number $N_{CS}$ is closely related to the gauge invariant Chern-Pontryagin CP number $N$.  In FIG \ref{figperiodic}, we see that
quantum tunneling from one vacuum state $\left|n\right\rangle$ (at integer $n=\mu/\pi$) to another can happen via $N=\Delta N_{CS}$: $\left|n\right\rangle \to \left|n+N \right\rangle$ \cite{Belavin1975,Hooft1976,Hooft1976a}. 
In studying the dynamics of the theory, we are interested in the transition from one vacuum to another via quantum tunneling. One can always study the time-dependent field equations in Minkowski spacetime \cite{Christ1980,Farhi1995,Gould1995a}, but this approach is rather complicated and little progress has been made. An alternative approach to study these quantum tunneling transitions is to generalize the CS number to a continuous dynamical quantum variable. (This property has been implicitly assumed in the literature but never worked out explicitly until recently \cite{Tye2015}.)
To this goal, we need an action or Lagrangian for the CS variable. With the potential $V(\mu)$ already worked out, we need only to find the kinetic term for the CS variable;  for the canonical kinetic form, we need only to find the mass coefficient $m$ and then canonically quantize the CS variable \cite{Tye2015}. This approach allows us to explore the underlying physics in a direction not possible before.


It turns out that the choice of the continuous CS variable is gauge-dependent and measure dependent. A necessary condition is that this CS variable has to agree with the gauge-invariant (under small gauge transformation) CS number at half-integer values. From the geometric point of view, the choice is arbitrary when moved away from half-integer values. 
So we may consider $N_{CS}=\mu/\pi + F(\mu)$, where $F(\mu)=0$ at $\mu/\pi \in \mathbb{Z}/2$. Based on the discussion below, it seems that $F(\mu)=-\sin(2\mu)/2 \pi$ is a natural choice.
Fortunately, physics consideration do suggest a simpler choice, namely, $N_{CS}=\mu/\pi$, i.e., $F(\mu)=0$.
Starting with the Lagrangian ${\bf L}$, canonical quantization of the CS variable $\mu/\pi$ yields the one-dimensional Schr\"{o}dinger equation,
\begin{equation}
\label{Sch1}
 {\bf L} ={1\over 2} m \left(\frac{\partial Q}{\partial t}\right)^2- V(Q) \quad \rightarrow \quad  \left( - {1 \over 2m}{\partial ^2 \over \partial Q^2} +V(Q) \right) \Psi(Q) = E\Psi(Q).
\end{equation} 
where $Q=\mu/m_W$ is chosen to have the dimension of length (with units so $\hbar=c=1$) so it mimics a spatial coordinate in standard quantum mechanics. Recently, the mass has been determined to be $m=17.1$ TeV \cite{Tye2015}. In general, choosing a non-zero $F(\mu)$ results in a $Q$-dependent mass $m(Q)$, which may even diverge for some values of $Q$ \cite{Tye2015}.

Notice that both the CP number $N$ and the CS number $N_{CS}$ are functions of the gauge fields $A^a_{\mu}(x)$ only. Physically the Higgs field couples to the gauge fields through the equations of motion, so it is expected that the Higgs field also contains relevant topological information. Now, the electroweak phase transition and spontaneous symmetry breaking are driven by the Higgs field, which leads to mass generation for the gauge fields and the fermions.  In particular, the dynamics of the electroweak phase transition and the $(B+L)$-violation are essential to the generation of the matter-anti-matter asymmetry of our universe. So it should be most convenient to link the change of $(B+L)$ value directly to the Higgs field instead of via the gauge fields. To this end, we like to find the relevant topological invariants that are functions of the Higgs field only. 
In this paper, we review the winding number $W$ of the Higgs field ($\pi_3(S^3)$) and construct the appropriate Hopf invariant for the Higgs field ($\pi_3(S^2)$). 

We can define a Hopf invariant $H_A$ that takes only (half-)integer values ($\pi_3(S^2)=\mathbb{Z}/2$) and another Hopf variable $H$ that takes continuous values. It is well known that the Hopf invariant measures the linking of the $S^1$s fibered over the base $S^2$, which takes integer values. At half integers, we see that the $S^1$s touch at a point, the cross-over that signifies the change of the linking number. That is, the half-integer Hopf invariant quantity is also gauge-invariant. Generalizing the Hopf invariant to other values for the Hopf variable is clearly gauge-dependent. With an appropriate choice of gauge, we can relate it to $N$ and $N_{CS}$ as well as their generalizations to continuous values. We discuss how $H_A$ and $H$ differ. 
In summary, we identify the winding number $W$ of the Higgs field with $N_{CS}$, so (choosing $\left|n=0\right\rangle$ as the reference vacuum), we have
\begin{align}
\label{WCS}
-H_A=&W=N_{CS}=N={\Delta B}/3={\Delta L}/3 \\\nonumber
m_WQ&=\mu/\pi= H=W=N_{CS}=N
\end{align}
where the top formulae apply when the quantities take discrete integer values as ${\Delta B}$ and ${\Delta L}$ are the discrete changes of the baryon number and the lepton number, respectively. The bottom formula is when the topological quantities are extended to continuous dynamical variables, as $\mu$ is treated as a real continuous variable. 
In short, the variable $Q$ in Eq.(\ref{Sch1}) also stands for the winding number $W$ or the Hopf variable $H$.
In this note we focus only on the $SU(2)$ gauge fields and the Higgs doublet. We like to go over the reasoning on the above identification (\ref{WCS}).

The construction of the discrete Hopf invariant $H_A$ and the Hopf variable $H$ in this paper is new. To discuss the relation of its extension to the CP and CS  variables, we briefly review these topological variables for the sake of completeness, i.e., the presentation here can be found in the literature, though it may be somewhat scattered so putting the relevant parts together may be of use to some readers. The rest of this paper is organized as follows.
In Sec. 2, we give a brief review of the relevant part of the electroweak model, where the quarks and leptons sectors and the $U(1)$ gauge part are ignored. In Sec. 3, we present the non-trivial static solution and discuss the relation between the CP number and the CS number. In Sec. 4, we discuss the winding number and the Hopf invariant as well as their relation to the CS number. In Sec. 5, we compare the different gauge choices and discuss the issues. In Sec. 6, we explain the justification of the choice of the suitable CS variable, namely $N_{CS}=W=H=\mu/\pi$. In Sec 7, we make a number of comments. Some details are relegated to an appendix.

\section{Background}
  
Here we like to discuss a geometric aspect of the electroweak model that is relevant for quantum tunneling. In this study, the Abelian $U(1)$ gauge field plays only a peripheral role, so we focus on the Yang-Mills (non-Abelian) $SU(2)$ gauge potential $A_{\mu}(x)$ and the Higgs (complex scalar) doublet field $\Phi$ only. In terms of the $2 \times 2$ Pauli matrices $\sigma^a$, $a=1,2,3$, we have (setting $c=\hbar=1)$,

\begin{align}\label{lagran} 
 \mathcal{L} = -{1\over2}{\Tr}[F_{\mu\nu}F^{\mu\nu}]+{1\over 2} \left(D_{\mu}\Phi\right)^{\dagger}D^{\mu}\Phi -{\lambda \over4}\left( \Phi^{\dagger}\Phi -v^2 \right) ^2 
\end{align}
where $g$ is the gauge coupling with $\alpha_W=g^2/4 \pi=1/30$, and $\mu$, $\nu$ are spacetime indices, $\mu, \nu =0,1,2,3$. 
\begin{align}
   F_{\mu\nu}&=F^a_{\mu\nu}{\sigma^a \over 2} = \partial_{\mu}A_{\nu} - \partial_{\nu}A_{\mu}-ig\left[A_{\mu},A_{\nu}\right] ,\nonumber\\
   D_{\mu}\Phi &= \partial_{\mu}\Phi - i g A_{\mu}\Phi, 
\end{align}
where $A_{\mu}(x)=A_{\mu}^a \sigma^a/2$. (Note that the index $\mu$ is unrelated to the $\mu$ variable to be introduced.) Under a gauge transformation,
\begin{equation}
\label{gtrans}
 A_{\mu} \rightarrow U A_{\mu} U^{-1} +\frac{i}{g}U \partial_{\mu} U^{-1}, \quad \quad \Phi \rightarrow U\Phi 
\end{equation} 
where $U(x)$ is an arbitrary $2 \times 2$ unitary matrix, each of the 3 terms in the Langrangian density $\mathcal{L}$ (\ref{lagran}) is invariant as well as the following equations of motions,
\begin{align}
\label{eom}
 D_{\mu}D^{\mu} \Phi = -\lambda(\Phi^{\dagger}\Phi - v^2)\Phi , \quad (D^{\mu}F_{\mu\nu})^a = -\frac{i}{4}g \left[ \Phi^{\dagger} \sigma^a D_{\nu}\Phi - (D_{\nu}\Phi)^{\dagger}\sigma^a \Phi \right],
\end{align}
where $D_{\alpha}F_{\mu\nu} = \partial_{\alpha}F_{\mu\nu} - ig \left[A_{\alpha},F_{\mu\nu} \right]$.

After spontaneous symmetry breaking, $\Phi$ acquires a vacuum expectation value $\Phi =(0,v)^T$ where $v=246$ GeV, and the gauge bosons develop a mass $m_W=gv/2=80$ GeV while the Higgs Boson has mass $m_H=\sqrt{2\lambda} v=125$ GeV. Note that the Higgs potential in $\mathcal{L}$ (\ref{lagran}) is the only renormalizable potential one can write down and its parameters $\lambda$ and $v$ are fixed by data. If we allow a more general Higgs potential, some of the quantitative details will be changed, but not the geometric properties to be discussed in this note. 

Following the homotopy group property $\pi_3(SU(2))=\mathbb{Z}$, the instanton solutions in 4-dimensional Euclidean space ${\bf R}^4$ have   
\begin{align} \label{Kcurrent}
N &= \frac{g^2}{16 \pi^2} \int d^4x {\Tr} \left[F_{\mu \nu} \tilde{F}^{\mu \nu} \right] = \int d^4x \partial_{\mu} K^{\mu},  \nonumber \\
\tilde{F}^{\mu \nu} &= \frac{1}{2} \epsilon^{\mu \nu \rho \sigma}  F_{\rho \sigma} \nonumber \\
K^{\mu} &= \frac{g^2}{32\pi^2}\epsilon^{\mu\nu\rho\sigma}\left(F^a_{\nu\rho}A^a_{\sigma} -\frac{g}{3}\epsilon^{abc} A^a_{\nu}A^b_{\rho}A^c_{\sigma}\right).
 \end{align}
where the Chern-Pontryagin (CP) number $N$  takes only integer values.

Notice that we have also expressed the gauge-invariant ${\Tr} \left[F_{\mu \nu} \tilde{F}^{\mu \nu} \right]$ in terms of the Chern-Simons current $K^{\mu}$, which is, however, gauge-dependent. Let us consider some localized non-zero field strength $F_{\mu \nu}$. Using Stokes theorem, we can express 
\begin{equation}
\label{Kterm1}
N (t_0) = \int d^3x K^0 {\bigg |}^{t=t_0}_{t=-\infty}  + \int_{-\infty}^{t_0} \int_{S} \vec{K}\cdot d\vec{S}.
\end{equation}
where $S$ is the surface area as the radius $r \rightarrow \infty$. If we take  $t_0 \rightarrow +\infty$, we recover $N$ in Eq.(\ref{Kcurrent}), i.e., $N(t_0=+\infty)=N$. This expression is valid even at finite time $t_0$. 

If $\vec K$ decreases fast enough at large distances, then the last term in Eq.(\ref{Kterm1}) vanishes so the Chern-Simons number
\begin{equation}
N_{CS}(A) =N(t_0)=\int d^3x K^0 {\bigg |}_{t=t_0} 
\end{equation}
where we have also assumed for simplicity that $K^0=0$ at $t=-\infty$. 

Under a gauge transformation (\ref{gtrans}) (and in differential form notation),
\begin{equation}
\label{CSn}
N_{CS}(A^U)=N_{CS}(A)+\frac{ig}{8\pi^2} \int   \Tr d \big[U^{\dagger} (d U) A \big] + \frac{1}{24\pi^2} \int \Tr\big[ (U d U^{\dagger})^3 \big]
\end{equation}
For sufficiently regular $A_i$ and $U$, the second term integrates to zero and the last term is simply the winding number of $U$.

For pure gauge
$$A_{\mu} = \frac{i}{g}U\partial_{\mu}U^{\dagger}$$
$$K^{\mu}= \frac{2}{3} \epsilon^{\mu\nu\rho\sigma}{\rm Tr}\left[ U \partial_{\nu}U^{\dagger} U\partial_{\rho}U^{\dagger} U \partial_{\sigma}U^{\dagger} \right]$$
Writing $U$ in terms of the unit radial vector ${\hat r}={\vec r}/r$ and magnitude $\mu$, we have 
\begin{equation}
\label{Umu}
U= \exp(i \mu \Omega(r) \hat{r}\cdot\vec{\sigma}) = {\bf I} \cos \mu\Omega(r)  + i\hat{r}\cdot\vec{\sigma} \sin \mu\Omega(r) 
\end{equation}
and the last term in the CS number (\ref{CSn}) becomes
\begin{align} 
\label{WU}
W(U)=N_{CS}(A) &=\frac{1}{24\pi^2} \int \Tr\big[ (U d U^{\dagger})^3 \big]=\frac{1}{2\pi}\big(2\mu \Omega(r) - \sin 2\mu \Omega (r)\big)\Big|^{r=\infty}_{r=0} \nonumber\\
& = \frac{1}{2\pi}\big(2\mu - \sin 2\mu \big)
 \end{align}
where $\Omega (r)=0$ at $r=0$ and $\Omega (r) \rightarrow 1$ as $r \rightarrow \infty$. 
 Notice that this winding number $W(U)$ of $U$ takes half-integer values for half-integer multiples of $\mu/\pi$. That is, under certain conditions, the CS number can be extended from integer values to half-integer values.
 The above formula also suggests that an extension to continuous values is also possible.  Since this is gauge dependent, it is important to find a guideline in the choice of a physically useful generalization.


\section{Static Solutions}

\begin{figure}[h] 
 \begin{center}
  \includegraphics[scale=.45]{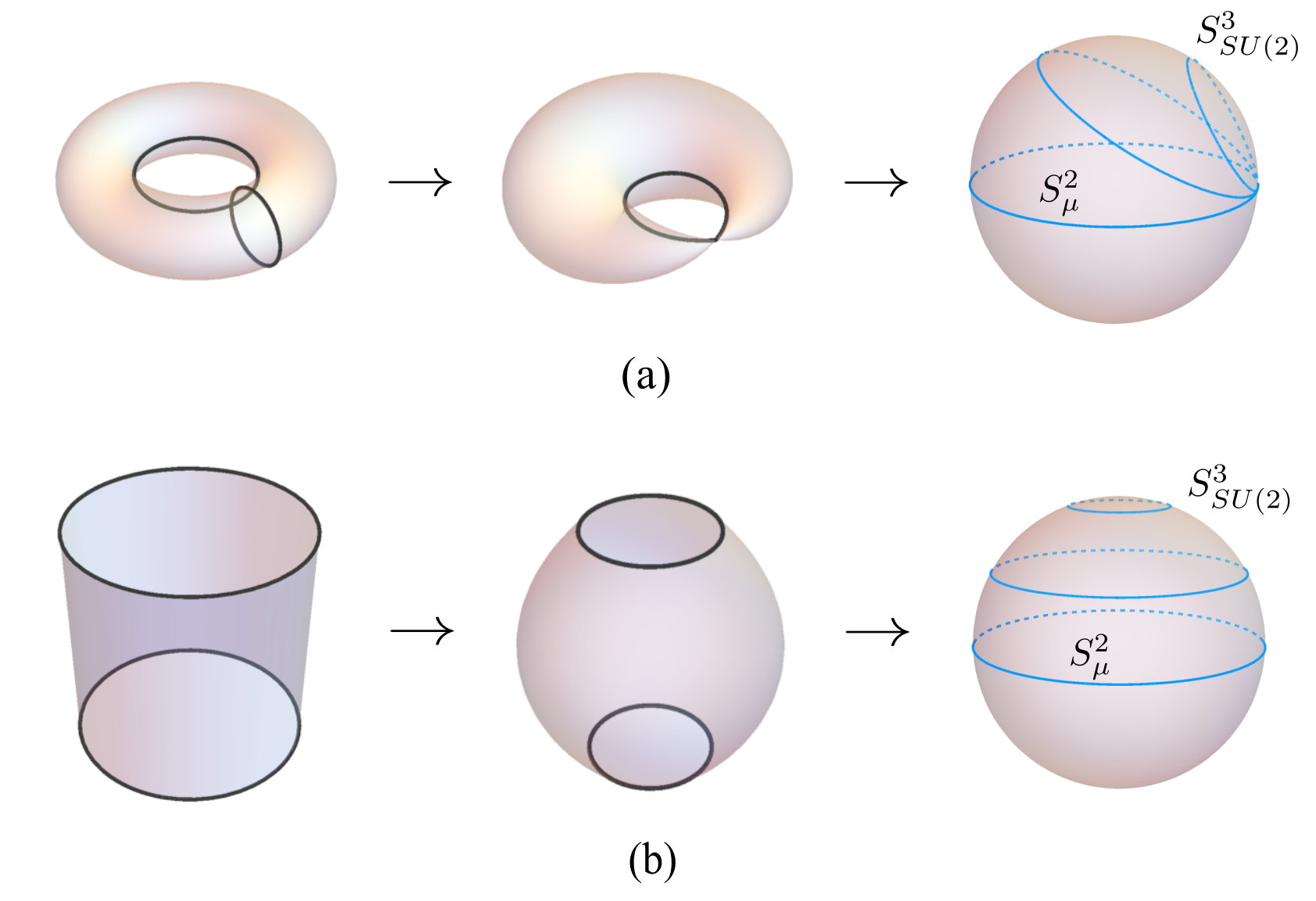}
   \caption{Here the 3-sphere $S^3_{SU(2)}$ is spanned by the point $p(\mu, \theta, \varphi)$ where $0 \le \mu < \pi$ and the usual polar angles $(\theta, \varphi)$ spans a 2-sphere. The sphaleron corresponds to $\mu=\pi/2$, when the 2-sphere attains maximum size (the equator in the sketch).  This unstable 2-sphere shrinks to a point at either vacuum ($\mu=0$ or $\pi$). There are two ways to cover the 3-sphere : (a) the one pole case, and (b) the 2 poles case.  (a) The top case illustrates the smash product $S^1 \wedge S^2=S^3$, where the black circles shrink to zero. This results in an $S^3$ with only one pole. (2) The bottom case illustrates how to get a $([0,\pi]\times S^2)/(\{0,S^2\}\cup \{\pi,S^2\}) =S^3$ with 2 poles, namely the north and the south poles. }  \label{FigS3}
   \end{center}
\end{figure}

Since we live in Minkowski space ${\bf R}^{3,1}$, it is convenient to introduce a corresponding number $n$ in Minkowski space. 
Let us now find static solutions to the equations of motion (\ref{eom}) of this model, limiting ourselves to the parameter range $0 \le \mu \le \pi$, so $\mu$ parametrizes the possible solutions. We shall choose the gauge where $A_0(x)=0$. 
One finds that there are 3 static solutions : \\
(1) $\mu=0$.  This vacuum solution is the simplest solution, where $A_j(x)=0$ and $\Phi(x)=v(0,1)^T$. \\
(2) $\mu=\pi$. It turns out that the vacuum is degenerate, that is, there is another vacuum where, at $\mu=\pi$, $A_j(x)=U^{\dagger}\partial_jU$ is a pure gauge (e.g., $U$ given in Eq.(\ref{Umu})). This has winding number $N=1$.  \\
(3) $\mu=\pi/2$.  In addition, the equations of motion yields a static solution at $\mu=\pi/2$. This ``sphaleron" solution is an unstable extremum (saddle point) solution of the model \cite{Manton1983}.  Let us discuss this solution in some detail.
We shall consider different gauge choices as $r \rightarrow \infty$ for the sphaleron solution: $\Phi$ takes either 
 a hedgehog (i.e., spherically symmetric) form (H), or is always aligned (A), that is, $\Phi$ asymptotes to the aligned/unitary vacuum, $\Phi \rightarrow (0,v)^T$. In covering the 3-sphere $S^3$, one can have either one pole or 2 poles (see FIG \ref{FigS3}). We shall consider the (1H), (2H) and (2A) gauges below. 
 
Consider the following ansatz in the (2H) gauge,
\begin{align}
\label{hedgehog}
   \Phi_H (\mu, r, \theta, \varphi) &=  v[1-h(r)] \begin{pmatrix}
   0 \\ \cos\mu   \end{pmatrix} + vh(r) U \begin{pmatrix} 0\\ 1 \end{pmatrix}, \nonumber\\
   \quad \hat{A}_i (\mu, r, \theta, \varphi) &= {i\over g} f(r) U \partial_i U^{\dagger},  \quad {\hat A}_0 =0,  \quad  U(\mu, r, \theta, \varphi)= \exp(i\mu\hat{r}\cdot \vec{\sigma} ), \nonumber\\
   \lim_{r\to 0}{f(r) \over r}&= h(0)=0, \quad f(\infty) = h(\infty) =1. \nonumber \\
\end{align}
In this ansatz, the Higgs field far from the origin goes from an aligned form to a hedgehog form as $\mu=0 \rightarrow \mu=\pi/2$ : $\Phi_H^{\infty}(\mu=0) = \big( \begin{smallmatrix}0 \\ v \end{smallmatrix} \big)$ changes to $\Phi_H^{\infty}(\mu=\pi/2) = i\hat r \cdot \vec{\sigma} \big( \begin{smallmatrix}0 \\ v \end{smallmatrix}\big)$.  The static equations of motion at $\mu=\pi/2$ become,
\begin{align}
\label{fhsol}
 r^2 f'' &= 2f(1-f)(1-2f) + m_W^2 r^2 h(f-1), \nonumber \\
 \left(r^2 h' \right)' &= 2h(1-f)^2 + \frac{1}{2} m_H^2 r^2 (h^2-1)h.
\end{align} 
Numerically, the 2 functions are well approximated by  $f(r) \approx 1-\mbox{sech}(1.154 m_W r)$ and $h(r) \approx \tanh (1.056 m_W r)$ \cite{Tye2015}. To summarize, we see that the system has no static solution unless $\mu/\pi \in \mathbb{Z}/2$.

\subsection{Recovering the Chern-Pontryagin Number}

For this static solution with $A_0=0$, we see that the gauge-invariant $\Tr F_{\mu \nu} \tilde{F}^{\mu \nu}=0$. To recover the CP number that links the 2 vacua, i.e., $\mu=0$ and $\mu=\pi$ (see FIG \ref{figperiodic}), we must introduce either a non-zero $A_0$ or a time-dependence somewhere.  It is simple to treat the parameter $\mu$ as a function of time $\mu (t)$, so that $\Tr F_{\mu \nu} \tilde{F}^{\mu \nu}\ne 0$. For $\mu(t = -\infty)=0$ and $\mu(t = +\infty)=\pi$, the integration over spacetime for $N$ (\ref{Kcurrent}) yields the CP number $N=1$, as expected. With $\mu(t)$ a function of time, we now can generalize the CP number $N$ to become a function of $t_0$ or $\mu (t_0)$. This also means that $\mu(t)$ becomes a dynamical variable. 

The sphaleron is the static solution where the potential barrier reaches its peak value $E_{sph}=9$ TeV  (see FIG \ref{figperiodic}). Classically, if we start with exactly this energy $E_{sph}$ and $\mu(t = -\infty)=0$, $\mu(t)$ will increase until it reaches the top and stay there at $\mu(t = \infty)=\pi/2$ (when $\mu$'s kinetic energy drops to exactly zero, hence static). In this case, we find that the CP number $N=1/2$. It is clear that there is no static solution for non-half-integer $\mu/\pi$. 

Given that the sphaleron solution exists, it is interesting to look for field configurations between the sphaleron and vacuum. However, there is no such solution to the static field equation of motion since we are not searching for the potential extrema in field space. It is appropriate to look for solution that minimize the potential under a constrained CS number. This was studied in Ref\cite{Akiba1988} and is reviewed in the appendix below. A simpler approach is to smoothly deform (parametrize) the field configuration from the sphaleron to vacuum, that is, one uses the above solution for $f(r)$ and $h(r)$ to extrapolate to other values of $\mu/\pi$ \cite{Manton1983}. (As we shall see, this turns out to be good enough for our discussion). 


In this sense, we can calculate the CP number due to the change in $\mu$. First note that $A_r^a=0$, 
\begin{align}
    \epsilon^{ijk}\tr \left[A_i\partial_j A_k \right] &= \epsilon^{irk}\tr \left[A_i \partial_r A_k \right] = \epsilon^{irk}\tr \left[ \mathcal{A}_i \mathcal{A}_k \right] f f' =0 , \nonumber  \\ \epsilon^{ijk}\epsilon_{abc}A_i^aA_j^bA_k^c &=\frac{1}{6} \det(A) =0\quad \Rightarrow K^0 = 0 , 
\end{align}
where $\mathcal{A}_i =\frac{i}{g}U\partial_i U^{-1} $. Therefore, we obtain 
\begin{equation}
\label{CPnum}
N = \frac{g^2}{16\pi^2} \int d^4x \Tr \left[ F_{\mu \nu} \tilde{F}^{\mu \nu} \right] = \int dt \int\vec{K}\cdot d\vec{S} =  \frac{2\mu - \sin 2\mu}{2\pi}f^2\bigg|^{r=\infty}_{r=0} =\frac{2\mu - \sin 2\mu}{2\pi},
\end{equation}
which reproduces the integers at $\mu=0, \pi$ and half-integer at $\mu=\pi/2$. Note that the CP number is gauge-invariant for $\mu/\pi \in \mathbb{Z}/2$. Notice also that $\mu(t)$ may be treated as a function of either Euclidean or Minkowski time.


\subsection{A Different Gauge}

In order to express the Lagrangian (\ref{lagran}) in terms the dynamical variable $\mu(t)$, one has to treat the gauge problem carefully. When time dependence is introduced to $A_{\mu}$ and $\Phi$, $A_0$ is in general non-zero and it is determined by Gauss law. Physically, the role of $A_0$ is to gauge away the non-physical rotations. For instance, the fields $A_i$ and $\Phi$ in ansatz (\ref{hedgehog}) are rotating as $\mu$ varies even far from the sphaleron. In the following, we would like to work in another ``gauge" and stay with $A_0=0$. This still does not minimize the kinetic term but has little effect in the phenomenology discussed in \cite{Tye2015}.

Let us switch to another ansatz where the Higgs field always stays in the unitary (i.e, aligned) vacuum far from the origin (i.e., the (2A) gauge), 
\begin{align}
\label{unitary}
   \Phi_A (\mu, r, \theta, \varphi) &= \tilde{U}^{\dagger} \Phi_H,\quad  A_i(\mu, r, \theta, \varphi) = \tilde{U}^{\dagger} \hat{A}_i \tilde{U} +{i\over g} \tilde{U}^{\dagger} \partial_i \tilde{U} ,  \quad A_0 =0, \nonumber\\
 \tilde{U}(\mu, r, \theta, \varphi) &= \exp(i\mu\Omega(r)\hat{r}\cdot \vec{\sigma} ), \quad  \Omega(0)=0, \quad \Omega(r \to \infty) \to 1 \, \mbox{exponentially}.
\end{align}
Note that $\Omega(r)$ can be any function satisfying the above condition since it gives the same energy functional as the hedgehog ansatz. It is because the above ansatz is a gauge transformation of the hedgehog ansatz in the static setup. If one wants to work in the gauge $rA_r=0$, we can go to a limit where $\Omega(r)$ is a step function of $r$ such that $\tilde U=U$ except at $r=0$. 
However, when $\mu$ is promoted to a function of time, $\mu(t)$, we have to take extra care in $\int \tr F\wedge F$ since this ansatz is not really a gauge transformation, 
\begin{equation}
  \frac{g^2}{16\pi^2} \int d^4x \Tr \left[ F_{\mu \nu} \tilde{F}^{\mu \nu} \right] = \frac{2\mu-\sin 2\mu}{2\pi} \left(f^2 +2(f-f^2)\Omega \right)\big| ^{r=\infty}_{r=0}= \frac{2\mu-\sin 2\mu}{2\pi}.
\end{equation}
It turns out that it gives the same $N$ as in the hedgehog ansatz due the the boundary condition of $f(r)$ (\ref{hedgehog}). We can also find the same result by using the CS form.
 At the spatial infinities, the ansatz always stays at the aligned/unitary vacuum $\Phi^T=(0,v)^T$. One can calculate the CS number of the ansatz, 
\begin{align} \label{2ACS}
  N_{CS} = \int d^3x K^0 &= \frac{2 \mu \Omega-\sin (2 \mu  \Omega) +4 f\sin \mu \sin (\mu  \Omega ) \sin (\mu -\mu  \Omega )}{2 \pi } \bigg|^{r=\infty}_{r=0} \nonumber \\
  &=\frac{2\mu - \sin 2\mu}{2\pi}.
\end{align}
Since $\vec{K}=0$ in the static gauge, we obtain $N=N_{CS}$. 

Note that we can start with the (2H) gauge and obtain the same result using (\ref{CSn}). Since $K^0=0$ in the 2H gauge and the second term in (\ref{CSn}) vanishes as a total derivative due to the regularity of $A$, $\int d^3x K^0 =\frac{1}{24\pi^2}\tr \int(\tilde{U}d\tilde{U}^{-1})^3$. The ansatz is in the unitary vacuum at spatial infinities, $A_i(|x|\to \infty) \to 0$ exponentially due to $\Omega(r)$, so one concludes that, 
\begin{equation}
  \int\vec{K}\cdot d\vec{S}=0 \quad \Rightarrow \quad N = \frac{g^2}{16\pi^2} \int d^4x \Tr \left[ F_{\mu \nu} \tilde{F}^{\mu \nu} \right] = \int d^3x K^0 \bigg|^{t=t_0}_{t=-\infty} =  \frac{2\mu - \sin 2\mu}{2\pi},
\end{equation} 
if we set the CS number density $K^0$ of the past reference vacuum to be zero.

\section{Topology in the Higgs Field}

Both the CP number $N$ and the CS number $N_{CS}$ as well as their generalizations to continuous values are functions of the gauge fields $A_{\mu}(x)$ only; they contain geometric information about the gauge fields.   It is interesting to see what geometric information is contained in the Higgs field. This is of great practical interest as phenomena such as finite temperature effects on the electroweak phase transition and spontaneous symmetry breaking and mass generation are all encoded in the Higgs field and these features play a crucial role in the matter-anti-matter generation via the sphaleron in the early universe.

On one hand, the equations of motion couple the gauge fields to the Higgs field, so we expect that some geometric properties of the gauge field will be transported to the Higgs field. On the other hand, there are $3 \times 4=12$ gauge fields $A_{\mu}^a$ while only 2 complex scalar fields in the Higgs doublet. In terms of the degrees of freedom, there are 6 degrees of freedom in $A_{\mu}^a$ while only 4 degrees of freedom in the Higgs field; so it is interesting to see how much of the geometric properties can be realized in the Higgs field alone. Here we like to consider the topological properties of the Higgs field and compare them to that of the gauge fields. In particular, we review the winding number $W$ coming from $\pi_3(S^3)$ of the Higgs field. We also construct the appropriate Hopf invariant from $\pi_3(S^2)$ of the Higgs field and relate it to $N$ and $N_{CS}$. 

\subsection{$\pi_3(S^3)$ in the Higgs Field}

 If we write the Higgs field as $\Phi = vh(\vec{x}) \mathcal{U}(\vec{x}) \xi$, where $h(\vec{x})$ is the magnitude of the Higgs field, $\mathcal{U}(\vec{x})$ is a $SU(2)$ unitary matrix and $\xi$ is a constant spinor. This complex doublet may be expressed as a 4 real component vector $\Phi$ to yield a 4-component unit vector $\hat{\Phi} =\Phi/|\Phi|$, i.e., $\hat{\Phi}^{\dagger} \hat{\Phi}=1$, which spans a 3-sphere $S^3$.   Then $\hat{\Phi} = \mathcal{U} \xi $ is a map $\hat{\Phi} : \mathbb{R}^3 \to S^3 $. If $\hat{\Phi}(x\to \infty)$ goes to a constant spinor, one can stereographically project $\mathbb{R}^3 \cup \{\infty\}$ to  $S^3$. In this sense, the map $\hat{\Phi}$ falls into the homotopy class of $\pi_3(S^3)$. By choosing $\xi = \big(\begin{smallmatrix}0 \\ 1 \end{smallmatrix} \big)$, $\mathcal{U}$ can be constructed from $\hat{\Phi}$ by, 

\begin{equation}\label{UPhi}
 \cal{U} = \begin{pmatrix}
   \hat{\Phi}_2^*  & \hat{\Phi}_1 \\
   -\hat{\Phi}_1^* & \hat{\Phi}_2
 \end{pmatrix}.
\end{equation}
We can obtain the winding number,  $$W(\mathcal{U}) = \frac{1}{24\pi^2}\int \tr(\mathcal{U}d\mathcal{U}^{-1})^3$$ 

\subsection{Hopf Mapping $\pi_3(S^2)$}

Given any unit spinor $\hat{\Phi}$, $\hat{\Phi}^{\dagger}\hat{\Phi}=1$, one can construct a Hopf mapping, 
\begin{equation}
  \eta: S^3_{\hat\Phi} \to S^2, \quad \hat{\Phi} \mapsto \hat{n}=\hat{\Phi}^{\dagger}\vec{\sigma}\hat{\Phi}.
\end{equation}
where $\hat{n} = \hat{\Phi}^{\dagger}\vec{\sigma}\hat{\Phi}$ is a 3-component unit vector, i.e., $\hat{n} \cdot \hat{n} =1$ due to the completeness relation $\vec{\sigma}_{\alpha \beta}\cdot\vec{\sigma}_{\gamma \delta}= 2\delta_{\alpha\delta}\delta_{\gamma\beta} - \delta_{\alpha\beta}\delta_{\gamma\delta}$. So $\hat{n}$ generates a $S^2$ and Hopf fibration has $S^2$ as base with $S^1$ as fiber, where the circles link once for Hopf invariant $H=1$.

What we are interested in is the composite map $\eta\circ \hat{\Phi}: \mathbb{R}^3\cup \{\infty\} \to S^2$, which falls into $\pi_3(S^2)$ when the spatial infinities are identified. In the following we just use $\hat{n}(\vec{x})$ to represent this function. The Hopf curvature is given by,
\begin{equation}
 b^i = \frac{1}{2} \epsilon ^{ijk}\epsilon_{abc} n^a \partial_j n^b \partial_k n^c = \epsilon ^{ijk}\partial_j a_k, \quad a_i = -2i \hat{\Phi}^{\dagger}\partial_i \hat{\Phi},
\end{equation}
where the second equality comes from the completeness relation. For instance, in (2A) gauge, we have
\begin{equation}
\begin{pmatrix}a_r \\ a_{\theta} \\ a_{\varphi} \end{pmatrix} = \begin{pmatrix} 2\mu \Omega' \cos \theta  -\frac{h'\cos \theta  \sin 2 \mu }{ \left(h^2 \sin ^2\mu +\cos ^2\mu \right)} \\\frac{\sin \theta  \left[\sin (2 \mu  \Omega ) \left(h^2 \sin ^2\mu -\cos ^2\mu \right)+h \sin 2 \mu  \cos (2 \mu  \Omega )\right]}{ \left(h^2 \sin ^2\mu +\cos ^2\mu \right)} \\ -2\frac{\sin ^2\theta  (h \sin \mu  \cos (\mu  \Omega )-\cos \mu  \sin (\mu  \Omega ))^2}{h^2 \sin ^2\mu +\cos ^2\mu } \end{pmatrix}
\end{equation}
The Hopf invariant is given by \cite{Whitehead1947}
\begin{equation}
  H = \frac{1}{16\pi^2}\int a \wedge da,
\end{equation}
where  $a = a_i dx^i$ is a one form. It can be shown that this is equivalent to $W(\mathcal{U})$ \cite{Jackiw2000}. We first define, 
\begin{equation}
  \mathcal{A}_i^a \frac{\sigma^a}{2}=  i\mathcal{U} \partial_i\mathcal{ U}^{-1},   \quad  \hat{m} = \xi^{\dagger}\vec{\sigma}\xi , \quad \mbox{such that}\quad  a_i =  \mathcal{A}_i^a \hat{m}^a.
\end{equation}
We also have $\epsilon^{ijk}\partial_j \mathcal{A}_k^a = -\frac{1}{2} \epsilon^{ijk}\epsilon_{abc}\mathcal{A}_j^b\mathcal{A}_k^c$ since $\mathcal{A}_i^a$ is a pure gauge. Hence,
\begin{align}\label{WHrelation}
 H &= \frac{1}{16\pi^2}\int d^3 r \epsilon^{ijk} a_i \partial_j a_k \nonumber \\
   & = \frac{1}{32\pi^2} \hat{m}^a \hat{m}^b \int d^3r \epsilon^{ijk}\epsilon_{bcd}\mathcal{A}_i^a\mathcal{A}_j^c\mathcal{A}_k^d \nonumber \\
   & = -\frac{1}{96\pi^2} \int d^3 r \epsilon^{ijk}\epsilon_{abc}\mathcal{A}_i^a\mathcal{A}_j^b\mathcal{A}_k^c \nonumber \\
   &=W(\mathcal{U}).
\end{align}
Therefore the Higgs field also gives the same winding number under this ansatz. Following from (\ref{WU}), the winding number can be read off from the boundary conditions. 

Let us find the Hopf invariant $H_A$ for $\Phi_A$ (\ref{unitary}).  Introducing $\omega (\mu,r)$ and recalling $\mathcal{U}$ (\ref{UPhi}), $\mathcal{U}_A$ can be expressed in term of $\omega (\mu,r)$, $\mathcal{U}_A=\exp\left(i \omega (\mu,r)\hat r \cdot \vec \sigma\right)$. We then have $H_A=W(\mathcal{U}_A)= \frac{1}{2\pi} (2\omega(\mu,r)-\sin(2\omega(\mu,r)))\big|^{r=\infty}_{r=0}$. The result is shown in the table below. 

\begin{center}
  \begin{tabular}{c || c| c |c}
          \multicolumn{4}{c}{Aligned/Unitary  (2A) Gauge}   \\
  $\mu$  & ${\hat \Phi}_A(r\to 0)$ & ${\hat \Phi}_A(r=\infty)$ &  $H_A$ \\  \hline
  $\left[0,\frac{\pi}{2}\right)$  &  $\scriptstyle{\begin{bmatrix}0 \\ 1 \end{bmatrix}}$ &  $ \scriptstyle{\begin{bmatrix}0 \\ 1 \end{bmatrix}}$    &    0     \\
   $\frac{\pi}{2}$ &  $\exp\left(i\frac{\pi}{2}\hat r \cdot \vec \sigma\right)\scriptstyle{\begin{bmatrix}0 \\ 1 \end{bmatrix}}$ & $ \scriptstyle{\begin{bmatrix}0 \\ 1 \end{bmatrix}}$ &  $-\frac{1}{2}$ \\
    $\left(\frac{\pi}{2},\pi\right]$  &    $\exp(i\pi\hat r \cdot \vec \sigma)\scriptstyle{\begin{bmatrix}0 \\ 1 \end{bmatrix}}$    &   $ \scriptstyle{\begin{bmatrix}0 \\ 1 \end{bmatrix}}$  &  $-1$ \\   
  \end{tabular}
\end{center}
We see that $H_A$ in the above (2A) ansatz (\ref{unitary}) jumps from 0 to -1 when $\mu$ passes through $\pi/2$. 

\begin{figure}[h]
 \begin{center}
  \includegraphics[scale=.5]{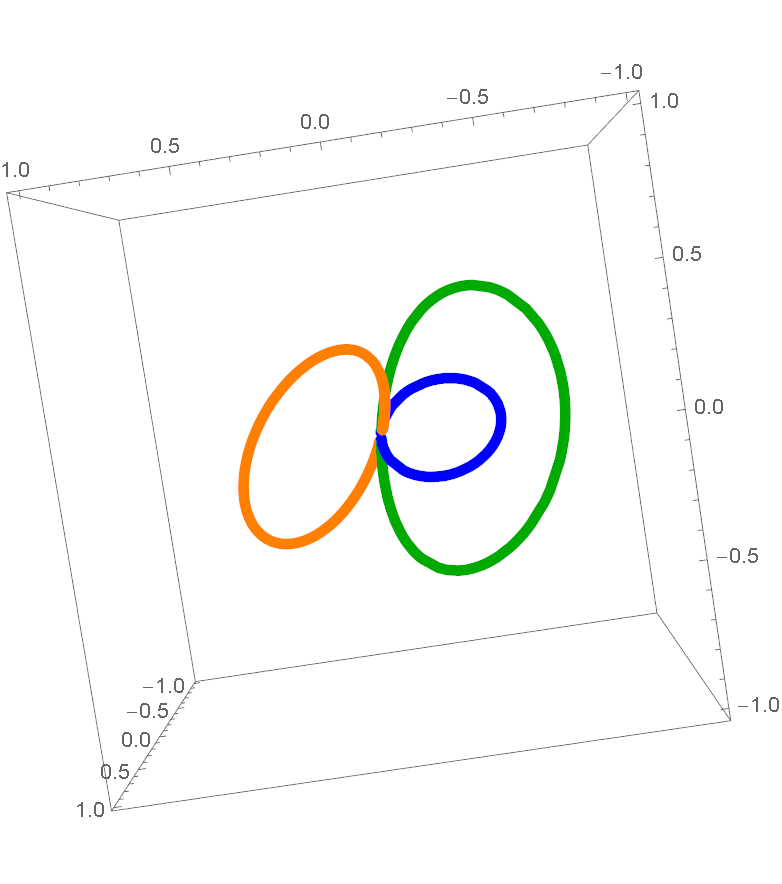}
  \caption{Illustration of the $\pi_3(S^2)$ mapping $\eta \circ \hat{\Phi}$ with Hopf invariant $H_A=-1/2$. The radial direction in $R^3\cup\{\infty\}$ is campactified in the plot such that $r=1$ corresponds to $\{\infty\}$. Each colored line corresponds to a circle $S^1$ fibered over a point in the base $S^2$. The 3 loops touch/intersect at the origin. For $0 \le \mu/\pi < 1/2$, $H_A=0$, when the 3 lines neither touch nor link. For $1/2 < \mu/\pi \le1$, $H_A=-1$, when any two loops link once but do not touch.}\label{linkhalf}
 \end{center}
\end{figure}

At the sphaleron, it gives $H_A=-1/2$ if we exert some care as $r \rightarrow 0$ since there is a coordinate singularity at the origin. For $\mu<\pi/2$, $\eta\circ \hat{\Phi}_A$ does not map $S^3$ to the full $S^2$ and the linking number is $H_A=0$. At $\mu=\pi/2$, $\hat{\Phi}_A$ maps $R^3\cup\{\infty\}$ to half of $S^3_{\hat\Phi}$ and $\eta$ further maps this half to $S^2$. The preimage $(\eta\circ \hat{\Phi}_A)^{-1}$ of points in $S^2$ are $S^1$ loops in $R^3\cup\{\infty\}$ passing through the origin, which indicates that the loops intersect (or touch, see FIG \ref{linkhalf}). Thus the linking number is $-1/2$. Their intersection is the jump from no linking to linking. For $\mu > \pi/2$, the preimage of any point in $S^2$ becomes a complete loop and all loops are linked once to each other without intersection. So the linking number is $H_A=-1$.
For our purpose, we may choose the Hopf invariant to be $H=-H_A$.

However, if we want to find a Hopf variable that matches the CP or CS continuous variable, we have to extend the above $H_A$ that takes discrete values to a continuous variable. To accomplish this, let us first go back to the (2H) ansatz (\ref{hedgehog}). Again, we can calculate the Hopf invariant $H_H$ using $\Phi_H$ (\ref{hedgehog}) and Eq.(\ref{WHrelation}); the result is shown in the table below. 

\begin{center}
  \begin{tabular}{c || c| c |c}
          \multicolumn{4}{c}{Hedgehog   (2H) Gauge }    \\
  $\mu$  & ${\hat \Phi}_H(r\to 0)$ & ${\hat \Phi}_H(r=\infty)$ &  $H_H$ \\  \hline
  $\left[0,\frac{\pi}{2}\right)$ &  $\scriptstyle{\begin{bmatrix}0 \\ 1 \end{bmatrix}}$  &  $ \exp(i\mu\hat r \cdot \vec \sigma)\scriptstyle{\begin{bmatrix}0 \\ 1 \end{bmatrix}}$    &    $\frac{2\mu -\sin 2\mu}{2\pi}$ \\
   $\frac{\pi}{2}$ &  $\exp\left(i\frac{\pi}{2}\hat r \cdot \vec \sigma\right)\scriptstyle{\begin{bmatrix}0 \\ 1 \end{bmatrix}}$ &  $\exp\left(i\frac{\pi}{2}\hat r \cdot \vec \sigma\right)\scriptstyle{\begin{bmatrix}0 \\ 1 \end{bmatrix}}$ &  0 \\
   $\left(\frac{\pi}{2},\pi\right]$  &     $\exp(i\pi\hat r \cdot \vec \sigma)\scriptstyle{\begin{bmatrix}0 \\ 1 \end{bmatrix}}$    &   $\exp(i\mu\hat r \cdot \vec \sigma)\scriptstyle{\begin{bmatrix}0 \\ 1 \end{bmatrix}}$   &  $\frac{2\mu -\sin 2\mu}{2\pi}-1$ \\   
  \end{tabular}
\end{center}

Here $H_H$ jumps from $({2\mu -\sin 2\mu})/{2\pi}$ to $({2\mu -\sin 2\mu})/{2\pi} - 1$ when $\mu$ passes $\pi/2$.
It is clear that the Hopf invariant is gauge dependent.
 
 \begin{figure}[h]
 \begin{center}
  \includegraphics[scale=.65]{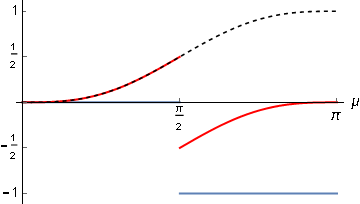}
  \caption{The Hopf invariant versus $\mu$. The red curve (with a jump at $\mu=\pi/2$) shows the Hopf invariant $H_H$ for the (2H) gauge. $H_A$ of the (2A) gauge is shown as the blue horizontal lines. The dash curve shows their difference, which is the appropriate Hopf invariant $H=H_H-H_A $ (\ref{Hsum}). }  \label{FigH}
 \end{center}
\end{figure}

\begin{figure}[h]
 \begin{center}
  \includegraphics[scale=.45]{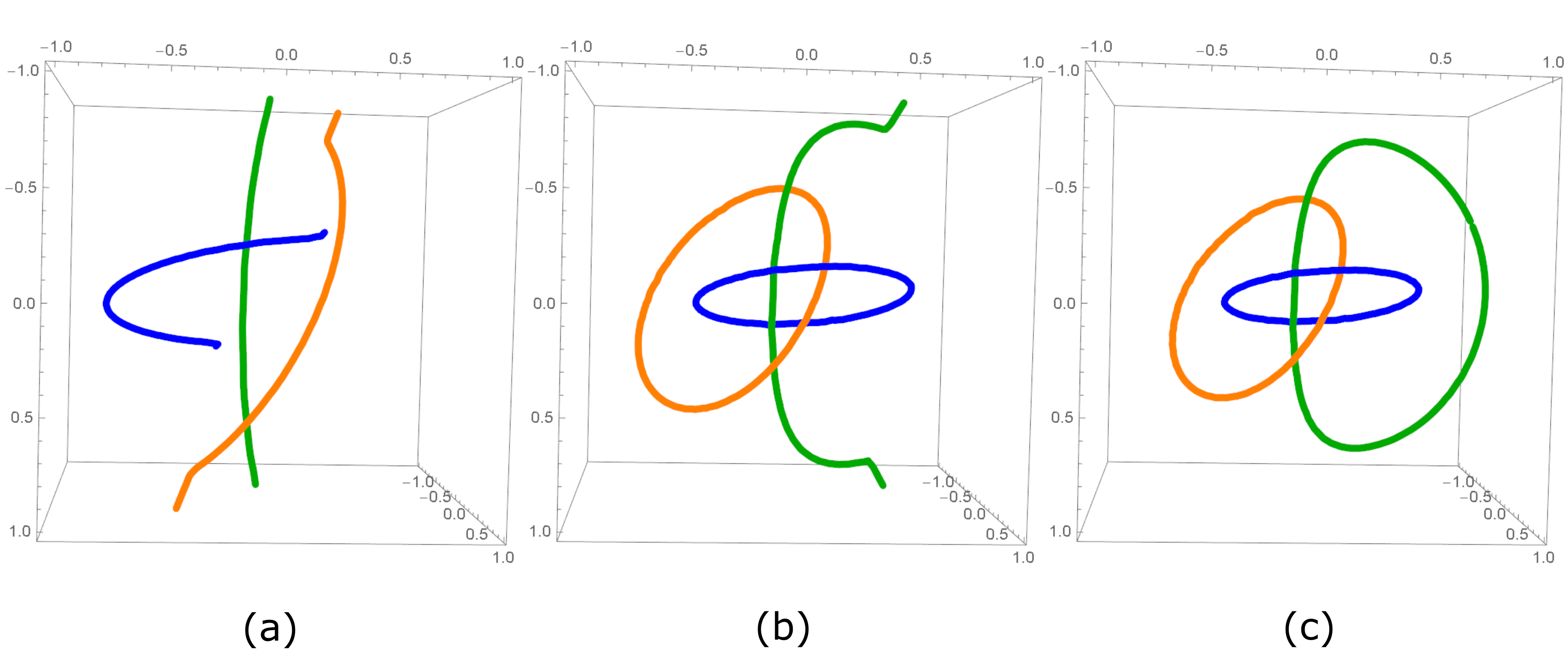}
  \caption{In the $\pi_3(S^2)$ mapping, every line/loop in $R^3\cup\{\infty\}$ is mapped to a point in the base $S^2$, which is not shown. The radial direction in the plots is  campactified such that $r=1$ corresponds to spatial $\{\infty\}$, and the $r=1$ surface is identified as a single point. So the two ends of each line are identified, i.e., it is actually a loop. 
 (a) At $H=1/2$, all loops touch at $\{\infty\}$; (b) at $H=0.9$, where some loops link but do not touch while the green loop still touches other loops at $\{\infty\}$;  (c) at $H=1$, when all loops link to each other once but do not touch. For $H<1/2$ (not shown), some loops disappear since the unit vector $\hat{n}$ does not span the full $S^2$ while the existing loops touch at $\{\infty\}$ but do not link.
  }  \label{linkpic}
 \end{center}
\end{figure}

To match the CS number $N_{CS}$ (\ref{2ACS}) obtained from the gauge fields, we construct the following Hopf invariant $H$ obtained from the Higgs field based on the above observations (see FIG \ref{FigH}), that is,
\begin{equation}\label{Hsum}
 H= -H_A +H_H = \frac{2\mu - \sin 2\mu}{2\pi},
\end{equation}
This construction can be achieved by patching the two gauges (hedgehog (2H) and aligned (2A)) to get the desired Hopf invariant $H$.  Given that the homotopic property is a group, we can always build a map $\hat{\Phi}$ from $\hat \Phi_A $ and $\hat\Phi_H$ such that it yields the desired $H$ (\ref{Hsum}). Such a construction is given by 
\begin{equation}
 \mathcal{U}=\mathcal{U}_A^{\dagger}\mathcal{U}_H, \quad \hat{\Phi} = \mathcal{U}\begin{pmatrix} 0\\ 1\end{pmatrix}= \begin{pmatrix}
\hat{\Phi}_{A2} \hat{\Phi}_{H1} -  \hat{\Phi}_{A1} \hat{\Phi}_{H2} \\ \hat{\Phi}^*_{A1} \hat{\Phi}_{H1} + \hat{\Phi}^*_{A2} \hat{\Phi}_{H2}
 \end{pmatrix}
\end{equation}
Due to (\ref{CSn}),
\begin{equation}
  \int\tr(\mathcal{U}d\mathcal{U}^{\dagger})^3 = \int\tr(\mathcal{U}_H d\mathcal{U}_H^{\dagger})^3 +  \int\tr(\mathcal{U}_A^{\dagger} d\mathcal{U}_A)^3 + 3i \int\tr d\left[\mathcal{U}_A^{\dagger} d\mathcal{U}_A\mathcal{U}_H d\mathcal{U}_H^{\dagger} \right].
\end{equation}
The last mixing term vanishes since $\mathcal{U}_A^{\dagger} d\mathcal{U}_A$ vanishes at the origin and infinity; so we arrive at the desired Hopf invariant,
\begin{equation}
   W(\mathcal{U}) = -W(\mathcal{U}_A)+W(\mathcal{U}_H) \quad \mbox{and} \quad H= -H_A +H_H.
\end{equation}
It can also be directly verified by checking the boundary conditions of $\hat{\Phi}$. We show the map $\eta \circ \hat{\Phi}$ for various $H$ in FIG \ref{linkpic}.

Instead of using both ans\"atze, one can also work in the following ansatz,
\begin{align}
\Phi = U\Phi_H, \quad U&= \exp\left(i\pi [\![\mu/\pi ]\!](\Omega(r)-1)\hat{r}\cdot\vec{\sigma} \right) \nonumber \\
[\![\mu/\pi ]\!]&= \frac{\lfloor \mu/\pi +1/2 \rfloor + \lceil \mu/\pi -1/2 \rceil}{2}
 \end{align}
where we have introduced the floor function, $\lfloor \mu/\pi +1/2 \rfloor = n \in \mathbb{Z}$ for $n \le \mu/\pi +1/2 < n+1$, and the ceiling function $\lceil \mu/\pi -1/2 \rceil = m \in \mathbb{Z}$ for $m-1 < \mu/\pi -1/2 \le m$.
This stair function jumps at $\mu/\pi\in \mathbb{Z} + 1/2$, when $[\![\mu/\pi ]\!]=\mu/\pi$. Such a jump essentially cancels the jump in $H_H$ at the same value of $\mu/\pi$ to give $H= ({2\mu-\sin2\mu})/{2\pi}$. 

In general, a Hopf invariant $H$ matches the CS number $N_{CS}$ only at the vacua (and sphaleron) since the above ansatz does not solve the static field equations of motion away from these configurations. This is also true in the AKY approach (see appendix). When it is neither sphaleron nor vacuum, the Hopf variable is a priori different from the CS variable in any  generic gauge choice. However, we can always define them to be matched, as is the case with H (\ref{Hsum}) and $N_{CS}$ (\ref{2ACS}). Such a matching is gauge-dependent, but it allows us to easily transport information contained in the gauge fields to the Higgs field.

In short, we now have
\begin{equation}
\label{WCS1}
H=W=N_{CS}=N=\mu/\pi +F(\mu), 
\end{equation}
where $F(\mu)=-\sin(2\mu)/{2\pi}$; so $F(\mu)= 0$ for   $\mu/\pi \in \mathbb{Z}/2$.


\section{Comparison of Ans\"atze}

Let us consider the (1H) gauge \cite{Manton1983},
\begin{align}
 \label{Mantoninfty}
   \Phi &= v[1-h(r)] \begin{pmatrix}
   0 \\ e^{-i\mu} \cos\mu   \end{pmatrix} + vh(r) U^{\infty} (\mu, \theta,\varphi) \begin{pmatrix} 0\\ 1 \end{pmatrix}, \nonumber\\
   A_i &= {i\over g} f(r) U^{\infty} \partial_i  (U^{\infty \dagger}) , \nonumber\\
 U^{\infty} &=
 \begin{pmatrix} 
    e^{i \mu}( \cos \mu - i \sin \mu \cos{\theta})  &    \sin \mu \sin{\theta} e^{i\varphi} \\
       -\sin \mu \sin{\theta} e^{-i\varphi} &  e^{-i\mu} (\cos \mu + i \sin \mu \cos{\theta})
\end{pmatrix},
\end{align}
where it differs from $\hat{\Phi}$, $\hat{A}_i$ in the (2H) gauge (\ref{hedgehog}) by the factor $e^{\pm i \mu}$. 

To see why $W(U)$ appears, let us introduce a 4-component vector ${\bf x}$,
$$U^{\infty \dagger}= x_4{\bf I} + i(x_i\sigma_i),$$ so that its determinant yields ${\bf x}\cdot{\bf x}=\sum x_i^2 =1$. We associate this unit vector ${\bf x}$ with a point $p(\mu, \theta, \varphi)$ in $S^3$, we may write
\begin{align}
\label{Manton2P}
  p (\mu, \theta, \varphi) 
   &=\begin{pmatrix}
    -\sin \mu \sin{\theta}\sin \varphi  \\ 
     -\sin \mu \sin{\theta} \cos \varphi \\ 
    - \sin \mu \cos \mu (1- \cos \theta ) \\ 
     \sin^2 \mu \cos \theta + \cos^2 \mu
\end{pmatrix}
\end{align}
where $(\theta, \varphi)$ are the standard polar coordinates for a sphere.
which spans the $S^3$ as shown in Fig. \ref{FigS3}(a). 
To obtain winding number $W(U)=1$, we take $\mu \in [0,\pi]$. So, instead of a Euclidean time, we stay in Minkowski time $t$ and introduce an angle $\mu (t)$, so $\mu(t=\infty)=0$ and $\mu(t = \infty)=\pi$. Notice that, at time $t=t_0$, when $\mu(t=t_0)=\pi/2$, we have $W(U)=1/2$. This corresponds to the maximum size $S^2$ generated by the polar coordinates $(\theta, \varphi)$. The construction can be seen as follows. We start with $S^2$ spanned by the standard polar coordinates $(\theta, \varphi)$. Introducing $\mu$ that spans $S^1$, we have the product $S^1 \times S^2$. By choosing $p(\mu, \theta, \varphi)$ such that $p(\mu=0, \pi)=(0,0,0,1)$ which is independent of $(\theta, \varphi)$ and $p(\theta=0)=(0,0,0,1)$ which is independent of $\mu$, we have the smash product of $S^1$ and $S^2$, $S^1 \wedge S^2 = S^3$. In this sense $U: S^3_{\{\mu,\theta,\varphi\}} \to S^3_{SU(2)}$ is characterized by $\pi_3(S^3)$.

The choice of $U$ is obviously not unique. For example, in the (2A) gauge (\ref{unitary}), we have
\begin{align}
\label{ours2}
  p (\mu, \theta, \varphi) &= \begin{pmatrix}
    \sin \mu \sin{\theta}\cos \varphi  \\ 
    \sin \mu \sin{\theta}\sin \varphi  \\
     \sin \mu \cos \theta \\
     \cos \mu 
\end{pmatrix}
\end{align}

This also spans the $S^3$ but in a different way, as shown in FIG \ref{FigS3}(b). In both cases, the choice of $\mu =\pi/2$ implies a maximum size for $S^2$. However, there is one pole ($p=(0,0,0,1)$) in the (1H) gauge while there are 2 poles ($p=(0,0,0,\pm1)$) in the coordinate (\ref{ours2}).


\section{Choice of a Chern-Simons Variable as a Generalization of the Chern-Pontryagin Number}

We like to find a continuous CS variable that agrees with the CP number at half-integers, but has a relatively simple form and transparent in revealing the underlying physics. Now a gauge transformation plus a redefinition allows us to go from one choice to another choice with the underlying physics remaining intact. However, recall the interaction between 2 electric charges in electrostatics. Although the physical description is in general gauge-dependent, the Coulomb potential (or force) in the Coulomb gauge clearly succinctly captures the picture. Similarly, the choice of the CS variable should be simple and best illuminate the underlying physics. This leads us to resort to physical reasoning to find a simple convenient choice.  
 
Recall that the complex Higgs doublet can be written as a 4 real component vector $\Phi$. Its unit vector $\hat \Phi = \Phi/ |\Phi|$ with coordinates $\{r, \mu, \theta, \varphi\}$ spanning a $S^3$. So we obtain $\pi_3(S^3)$. We introduce a winding number that equals unity ($W=1$) as $\mu$ spans $0 \rightarrow  \pi$. If $\mu$ spans only $0 \rightarrow  \pi/2$, we should get winding number equals to $W=1/2$ via symmetry (see FIG \ref{FigS3}). By continuity, as $\mu$ spans $0 \rightarrow  \mu_0$, the winding number $W$ can take any value $\mu_0/\pi$ plus a function $F(\mu_0)$ that vanishes at half-integer values of $\mu/\pi$. The simplest choice is to set $F(\mu_0)=0$, so the winding number is simply given by
 \begin{equation}
 \label{Wmupi}
W(\mu)=\frac{\mu}{\pi}.
\end{equation}
At first sight, based on the above analysis, a natural choice would be $F (\mu)=- \sin (2 \mu)/2\pi$, or
$$\hat W(\mu)= \frac{\mu}{\pi}  - \frac{\sin(2\mu)}{2 \pi},$$
where, at $\mu/\pi \in  \mathbb{Z}/2 $,
$$W=\hat W=N=N_{CS}=H \in \mathbb{Z}/2 .  $$
However, we like to give 2 reasons why $\hat W$ may not be a good choice. (Of course, any choice can be transformed to any other choice, so the end result is a matter of convenience and the transparency of the underlying physics.)  \\
(1)As shown in the appendix, a different way in extrapolating $\mu/\pi$ to non-half-integer values results in a different $F(\mu)$, so 
 the function $F(\mu)=- \sin (2 \mu)/2\pi$ is not unique. \\
(2) In treating $W(t)=\mu(t)/\pi$ as a dynamical variable, we have to find the kinetic term for it, i.e., the $\frac{M}{2} (\frac{\partial \mu}{\partial t})^2= M\dot{\mu}^2/2$ in 
\begin{equation} 
\label{LQ}
   {\bf L}=\int d^3x {\cal L} ={ 1\over 2} m {\dot Q}^2 - V\left(Q\right),
\end{equation}
where we rescale the winding number with the W-boson mass $m_W$ so $Q(t)=\pi W(t)/m_W =\mu(t)/m_W$ has the dimension of a length scale just like that of an ordinary coordinate. The Lagrangian density ${\cal L}$ (\ref{lagran}) has kinetic terms  for both $A_{\mu}(x)$ and $\Phi (x)$ that include the time derivatives. To obtain the kinetic terms for $W$, we start with  the static solutions for $A_{i}(\mu, r, \theta, \varphi)$ and $\Phi (\mu, r, \theta, \varphi)$ and then introduce time-dependence only via the time dependence in $\mu(t)$ or $W(t)$ : $A_{i}(\mu(t), r, \theta, \varphi)$ and $\Phi (\mu(t), r, \theta, \varphi)$. This is how we obtain the CP number (\ref{CPnum}). Following this approach, 
$$ \frac{\partial A_{i}}{\partial t} =   \frac{\partial A_{i}}{\partial \mu} \frac{\partial \mu}{\partial t}, 
\quad \quad  \frac{\partial \Phi}{\partial t} =   \frac{\partial \Phi}{\partial \mu} \frac{\partial \mu}{\partial t}. 
$$ so the kinetic terms in  ${\cal L}$ (\ref{lagran}) yield a kinetic term for $Q(t)$ (\ref{LQ}). In this approximation, one finds the mass to be a finite constant, $m= 17.1$ TeV, that is, the kinetic term has the canonical form \cite{Tye2015}. On the other hand, in the kinetic term for $\hat W(t)$, we obtain a mass ${\hat M}(\hat W(t))$ that depends on $\hat W(t)$ and diverges as $\hat W$ approaches vacuum values \cite{Tye2015}. 


\section{Discussion}

So far, we have barely mentioned the fermions that are present in the standard electroweak theory. We have 2 types of fermions in nature : quarks, which interact with the nuclear force, and leptons, which do not.  It turns out that a change of the CP number $N$ leads to a change in both the baryon number $B$ and the lepton number $L$ \cite{Hooft1976} (here we shall not worry about the fact that there may be 3 different lepton numbers 
), 
$$ \Delta B = \Delta L = 3N,$$
where the factor of $3$ is because nature has 3 families of quarks and leptons. The vacuum degeneracy implies that our universe is sitting in a specific vacuum labelled by $\left|n\right\rangle$ (see FIG \ref{figperiodic}).
For $N=1$, our universe moves from $\left|n\right\rangle \to \left|n+1 \right\rangle$ and there is an increase of 3 additional  baryons and 3 additional leptons in our universe.  That is, both $B$ and $L$ are not conserved (while $(B-L)$ is conserved). This non-conservation allows the possibility that matter-anti-matter asymmetry in our universe was generated via the sphaleron physics in the early universe during the electroweak phase transition.

During the electroweak phase transition, the vacuum expectation value of the Higgs field $\left\langle\Phi\right\rangle$ goes from zero to $v$. Yukawa couplings of the quarks and leptons to $\Phi$ generate the masses for the fermions, which are massless before the spontaneous symmetry breaking. By endowing the Higgs fields with geometric properties, one may hope to study the sphaleron dynamics and the generation of matter-anti-matter asymmetry by focusing on $\Phi$ and fermions only. 
This should simplify the analysis as the geometric properties of gauge fields are quite non-trivial at finite temperature and in phase transition.

\appendix
\section{A Different Functional Form} 

\begin{figure}[h]
 \begin{center}
  \includegraphics[scale=.6]{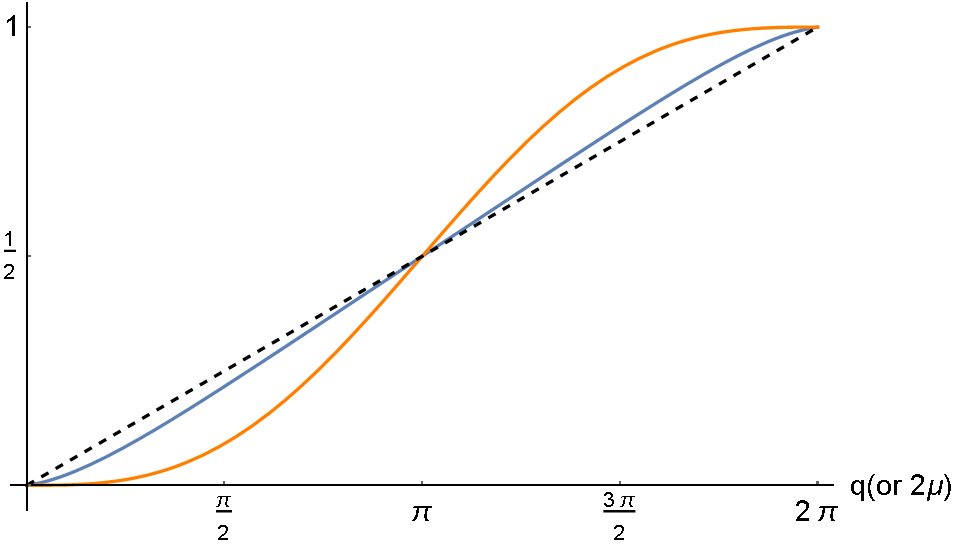}
  \caption{$N$ versus $2\mu$. Comparison of $N(\mu)=\mu/\pi - \sin(2\mu))/2\pi$ (\ref{CPnum}) (orange line) and  $N(q)$ in AKY (blue line). The dashed line is the linear line given by  $N(\mu)=\mu/\pi$. That is, $F(\mu)$ in AKY takes a different form (with smaller values) when compared to the $F(\mu)= - \sin(2\mu))/2\pi$ form in the paper.}  \label{FigAKY}
 \end{center}
\end{figure}

In the above analysis, we see repeatedly that the index value away from half-integers is always accompanied by the factor $\sin (2 \mu)/2\pi$. Here, we like to remind the reader that this needs not be the case by reviewing the approach due to Akiba, Kikuchi and Yanagida (AKY)\cite{Akiba1988}, where a different functional form emerges (shown in FIG \ref{FigAKY}).

Recall that the static solution exists only at an extremum of the energy functional, which happens at half-integer values of $\mu/\pi$. Away from them, there is no static solution. We expect the potential $V(\mu)$ to be a smooth function of $\mu$ (see FIG \ref{figperiodic}). Away from the extrema, no static solution exists for the obvious reason that $\mu(t)$ likes to evolve, implying only a time-dependent solution can be found. However, we still like to find the shape of the potential away from the extrema. To determine what the potential $V(\mu)$ looks like, one way is simply to  extrapolate from the extrema. Another way is to introduce a Lagrange multiplier (or chemical potential) to force for a static solution \cite{Akiba1988}. This approach starts with the most general spherically symmetric static ansatz,
\begin{align}\label{AKY}
  A_0^a &= \frac{1}{g} a_0(r)\hat{x}^a, \quad
  A_j^a = \frac{1}{g} \left[ a_1(r) \hat{x}_j\hat{x}_a + \frac{f_A(r)-1}{r}\epsilon_{jam}\hat{x}_m + \frac{f_B(r)}{r}( \delta_{ja} - \hat{x}_j\hat{x}_a) \right],  \nonumber \\
  \Phi & = \left( h(r) +  k(r)i \vec{\sigma}\cdot \hat{x} \right)\begin{pmatrix} 0\\ v
   \end{pmatrix},
\end{align}
and adds a Lagrange multiplier term $\eta (N[A] -n)$ to $\mathcal L$ such that the equations of motion have solution for any $n$. The CP number (after switching to unitary gauge) is given by
\begin{align}
  N= \int d^3x K^0 +\frac{q-\sin q}{2\pi} =  \frac{q}{2\pi} + \frac{1}{2\pi} \int dr {\Re}(i\chi^* \partial_r \chi),
\end{align}
where $q/2$ plays the analogous role as $\mu$ and a reference vacuum of zero CP number is chosen at past infinity. Here, $\chi = f_A+if_B$ and its integral in the last term is a periodic function that vanishes at integer values of $q/\pi$. However, it is quite different from the sinusoidal form we have been seeing. This CP number $N$ is shown in FIG \ref{FigAKY}. We made a gauge transformation of $ \Omega(r)= -q \lim_{\beta \rightarrow \infty} \tanh(\beta r)$ to the AKY boundary conditions to obtain our boundary conditions in order to preserve regularity at the origin. This changes $a_1(r)$ in (\ref{AKY}) to a $\delta$-function. This also changes the CP number and its effect is already explicitly included in the term $(q-\sin q)/2\pi$ while $\int d^3x K^0$ is safely calculated from the rest.

The corresponding Hopf invariant in the AKY ansatz is
\begin{equation}
 H= \frac{q -\sin q}{2\pi},
\end{equation}
with proper covering of gauge choices. This can be read off from the boundary conditions of $h(r)$ and $k(r)$ \cite{Akiba1988}.

 \vspace{5mm}
 
The work is supported by the CRF Grant HKUST4/CRF/13G and GRF 16305414 issued by the Research Grants Council (RGC) of Hong Kong.

\bibliographystyle{utphys}
\bibliography{References}

\end{document}